\pdfoutput=1
\documentclass[review]{elsarticle}

\usepackage{hyperref}
\hypersetup{
    colorlinks=true,      
    urlcolor=blue,
    linkcolor=black,
    citecolor = black}

\begin{document}

\begin{frontmatter}

\title{The energy spectrum of ultraheavy nuclei above 10$^{20}$ eV}

\author{Antonio Codino
}
\address{University of Perugia, Italy}




\begin{abstract}
A major feature of the energy spectrum of the cosmic radiation above 10$^{19}$ eV is the increasing fraction of heavy nuclei with respect to light nuclei. This fact, along with other simple assumptions, is adopted to calculate the energy spectrum of the cosmic radiation up to 2.4$\times$10$^{21}$ eV. The predicted spectrum maintains the index of 2.67 observed at lower energies which is the basic, known, empirical well-assessed feature of the physical mechanism accelerating cosmic rays in the Galaxy. Indeed above 10$^{19}$ eV the injection of nuclei is inhibited by some filter and this inhibition causes a staircase profile of the energy spectrum. It is argued that particle injection failure versus energy commences with protons, followed by Helium and then by other heavier nuclei up to Uranium. Around 7.5$\times$10$^{20}$ the cosmic radiation consists solely of nuclei heavier than Copper and the estimated intensity is 1.8$\times$10$^{-30}$ particles/GeV s sr m$^2$.
\end{abstract}


\end{frontmatter}

\section{Introduction}
In the year 2007 came to light the unexpected and severe result \cite{unger07}  that a large fraction of the cosmic radiation  above the ankle energy of  3.1$\times$10$^{18}$ eV consisted of  heavy nuclei and not only of proton and Helium. The outcome was confirmed by another two independent  methods of measurements \cite{cazon,unger09a} feasible with the unsurpassed Auger instrument. After the year 2013 the chemical composition resulting from   X$_{max}$ and  the width of the X$_{max}$   distribution observed by the Auger  Collaboration  has become  unmatchable  with  that  reported  in the period 2007-2011  as  argued in Section 3. Presently (2017) the preponderance of heavy nuclei above 5$\times$10$^{18}$ eV is based more on the X$_{max}$  measured by the Telescope Array (hereafter TA) detector  rather than on  recent data of the Auger Group.

The evolution of the chemical composition of the cosmic radiation toward heavy nuclei is of paramount importance since it entails the reorientation of some basic concepts in  Cosmic Ray Physics. One of these concepts is that  cosmic rays of maximum observed energies 10$^{19}$-3.0$\times$10$^{20}$ eV  do not come from  extragalactic  sources but are domestic, of galactic origin. Cosmological nuclei would be photodissociated and destroyed before harbouring in the Milky Way Galaxy and hypothetical cosmological protons  will be decelerated via photopion reactions,  ultimately  stranded  in local ambients without intercepting  terrestrial instruments.

According to this study the increasing fraction of heavy nuclei above 10$^{19}$ eV  observed by TA and Auger experiments (see figure \ref{fig_3}) will continue as energy ascends, becoming irresistible;  for example,  above the energy of  6.76$\times$10$^{20}$ eV the cosmic radiation is expected to consist only  of nuclei heavier than  Iron.  Nuclei are predicted to disappear from the cosmic-ray flux because the injection to the acceleration process is inhibited by a filtering process, or something equivalent to a filtering process,  operating  at the galactic sources.  The sieve  does initiate at the energy of  2.6$\times$10$^{19}$ eV \cite{codino17}. The Galactic sources are located  in the cold boundaries wrapping up  the H II regions embedded in the  O B star associations of the Milky Way Galaxy  (as it will be described in \textit{The rule governing cosmic ray abundances prior to acceleration} paper in preparation).
\section{The features of the energy spectrum above 10$^{19}$~eV}

\begin{figure}[h]
  \begin{center}
    \includegraphics[width=0.6\textwidth]{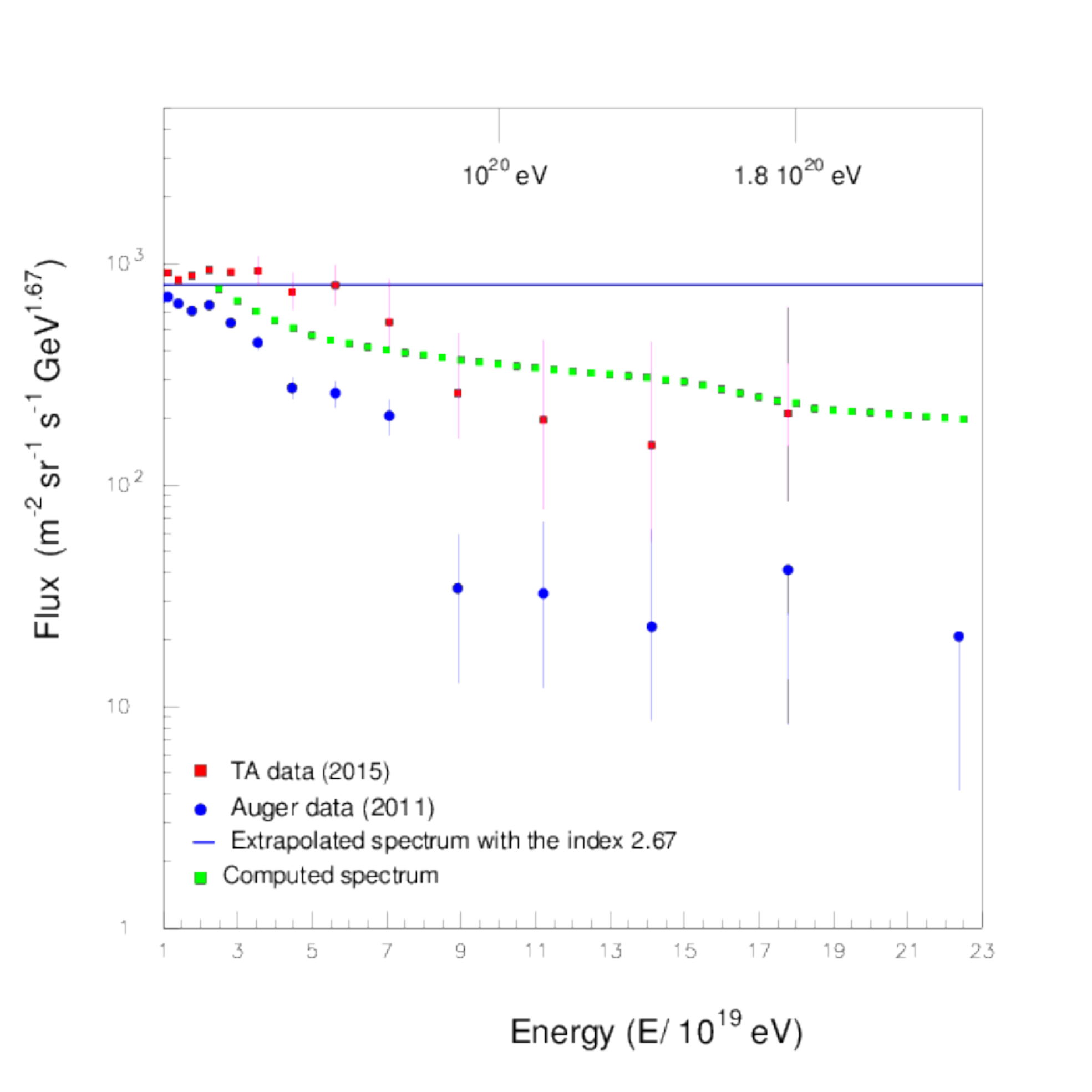}
    \caption{\label{fig_1}The computed energy spectrum  is represented by green squares in the range 10$^{19}$-2.3$\times$10$^{20}$ eV  which is the highest energy interval where experimental data are available.  The data  measured by Telescope  Array  (red squares) \cite{tinyakov}  and Auger Collaborations (blue dots) \cite{aab} are shown for comparison with the predicted spectrum. The  two Auger data points at the highest energies  are upper limits to the flux.}
\end{center}
\end{figure}

\begin{figure}[h]
  \begin{center}
    \includegraphics[width=0.6\textwidth]{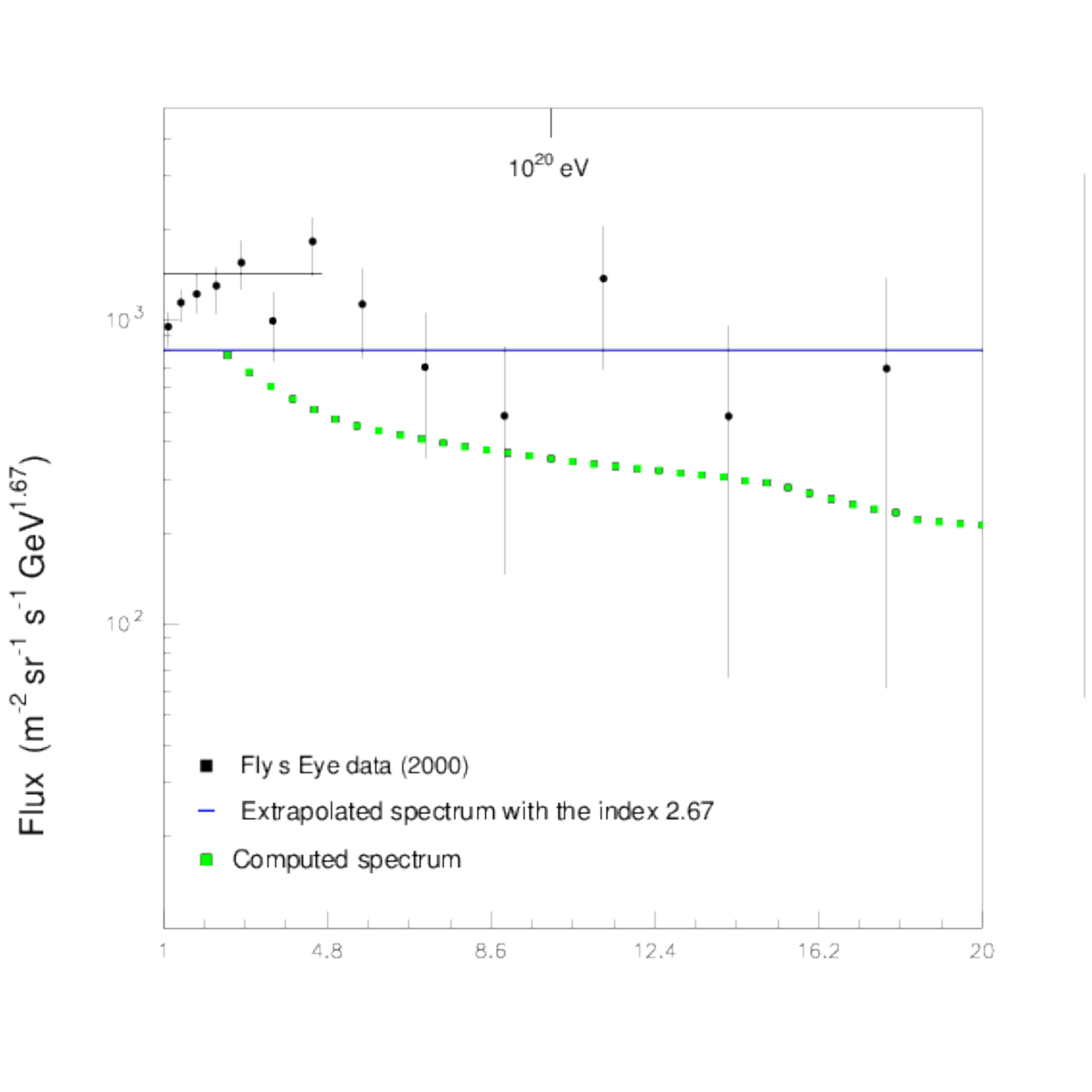}
    \caption{\label{fig_2}Comparison of  the computed spectrum (green squares)  with the flux data of the Fly's Eye Experiment  (black dots)  \cite{jui} in the same frame of fig. 1}
\end{center}
\end{figure}

The domestic origin of ultrahigh cosmic rays has been assumed in a recent calculation \cite{codino17} of the energy spectrum of the cosmic radiation in the interval  10$^{19}$-2.4$\times$10$^{21}$ eV.  The  part  of the computed spectrum where experimental data are available i.e. 10$^{19}$-3.0$\times$10$^{20}$ eV is shown in figure \ref{fig_1} (green squares)  along with the fluxes measured by Telescope Array \cite{tinyakov} and Auger experiments \cite{aab}. Figure \ref{fig_2} shows the fluxes measured by the Fly's Eye  \cite{jui}  experiment that took data in the period 1997-2006 with a final exposure of 4500 km$^2$/year sr and now dismantled and recycled.


The energy  spectrum in figures \ref{fig_1} and \ref{fig_2}  (green squares) shows a distinctive silhouette,  visible in the appropriate variables: the energy E  in a linear scale  and  the flux multiplied by  E$^{\gamma}$  which mitigates the steep fall of the spectrum with energy  (E is the particle energy and $\gamma$  the spectral index of 2.67). While a linear scale of energy to visualize the spectrum is rare,  the multiplication of the flux by   E$^{\gamma}$  with the desired  $\gamma$  is  routine.  Notice that a logarithmic scale in energy would have compressed the data points, belittling  the distinctive and  unique silhouette of the spectrum  in the interval  10$^{19}$-1.8$\times$10$^{20}$  eV  gleaned by the Auger apparatus \cite{aab}. The spectrum profile (green squares) is echoing the abundances of quiescent interstellar atoms at the sources prior to acceleration (see figure \ref{fig_3} of ref. \cite{codino17})  and the universal  index  $\gamma$  of  the Galactic Accelerator  (Part 3  of ref. \cite{codino15a}).

Calculation details are found in a previous paper \cite{codino17} and here only the basic tenets of the spectrum calculation are summarized: (1) the physical process accelerating cosmic rays takes place in the Galaxy and is not known; to designate the unknown acceleration process the term \textit{Galactic Accelerator} is used. Although the acceleration mechanism is unknown it has some identified, constrained features: (2) it delivers cosmic rays with a constant spectral index of  2.67  up to the energy of  2.4$\times$10$^{21}$ eV. (3) The acceleration process is not localized in any celestial bodies but it is ubiquitous in the Galactic volume.

The event suppression in the spectrum discovered by the HiRes Group in 2004 \cite{abbasi}  is interpreted as the  maximum energy attainable  by protons in the Galaxy. The physical mechanism causing the break is called \textit{LIGA} effect  (for \textit{Lack of particle Injection to the Galactic Accelerator}). All the important tenets above are nested in a reasoning (Section 3,  ref.  \cite{codino17})  leading to the predicted spectrum shown  in figure \ref{fig_1} and \ref{fig_2} .

     The  interpolation of the spectral break \cite{abbasi}   above  3$\times$10$^{19}$ eV via a power law with a single ultrasoft  slope, ($\approx$ 3.5-5.5)   has been   performed by  HiRes,  Auger and  TA  experiments  in the investigation of  the highest energy cosmic-ray events, a few dozens of events.   According to these three groups, as explicitly stated in many papers,  the break interpolation via an ultrasoft index proves the existence the hypothetical GZK effect. But this interpretation  plainly conflicts with the experimental data reported by the same experiments as described a few years ago \cite{codino13}.
\section{Empirical  basis  of the calculation of the energy spectrum}
The derivation of the computed spectrum (green squares in figure \ref{fig_1} and \ref{fig_2}) is based on the important assumption (A):  the chemical composition of the cosmic radiation evolves from light to heavy  in the range  5.0$\times$10$^{18}$- 10$^{20}$ eV.  This Section deals with the empirical foundation of this assertion (A)  which was omitted in the preceding  paper \cite{codino17}  due to its  small size.  From the Auger Group:  `` \dots as can be seen measurements favour a mixed composition.''   Michael Unger (2008)  \cite{unger09b}.

From the published results \cite{unger07,cazon,unger09a} of the Auger experiment in the years 2007-2012  emerges the picture that the chemical composition of the cosmic radiation in the interval   4.0$\times$10$^{18}$- 4.0$\times$10$^{19}$ eV  consists of a substantial fraction of intermediate and heavy nuclei. On the contrary, the TA Collaboration believes that,  in this energy range,  the cosmic radiation is dominated by protons as explicitly expressed in a number of papers. This last credence simply reiterates that of the HiRes  Group \cite{abbasi05a} that previously operated a florescence  detector  on the same geographical site (Dugway, Utah, North America,  39 Nord,  120 West).

\begin{figure}
  \begin{center}
    \includegraphics[width=0.6\textwidth]{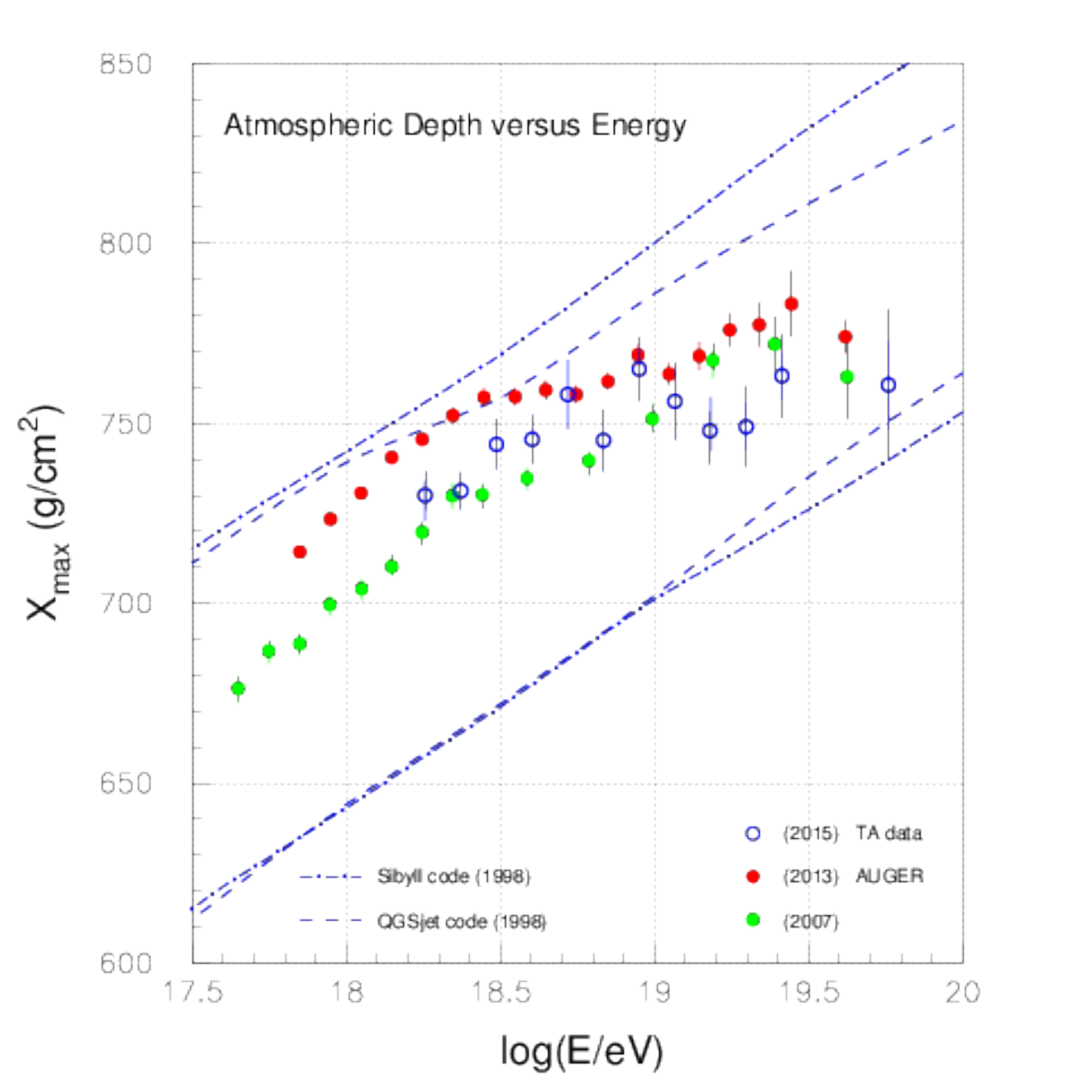}
    \caption{\label{fig_3}Atmospheric depth versus energy  measured by the Telescope Array \cite{fukushima,abbasi14} and Auger
Collaborations \cite{unger07,ahn}.  Theoretical depths  for  purely H and  Fe  cosmic nuclei   (blue lines called rails)  result from classical calculations \cite{heck}. }
\end{center}
\end{figure}

\begin{figure}
  \begin{center}
    \includegraphics[width=0.6\textwidth]{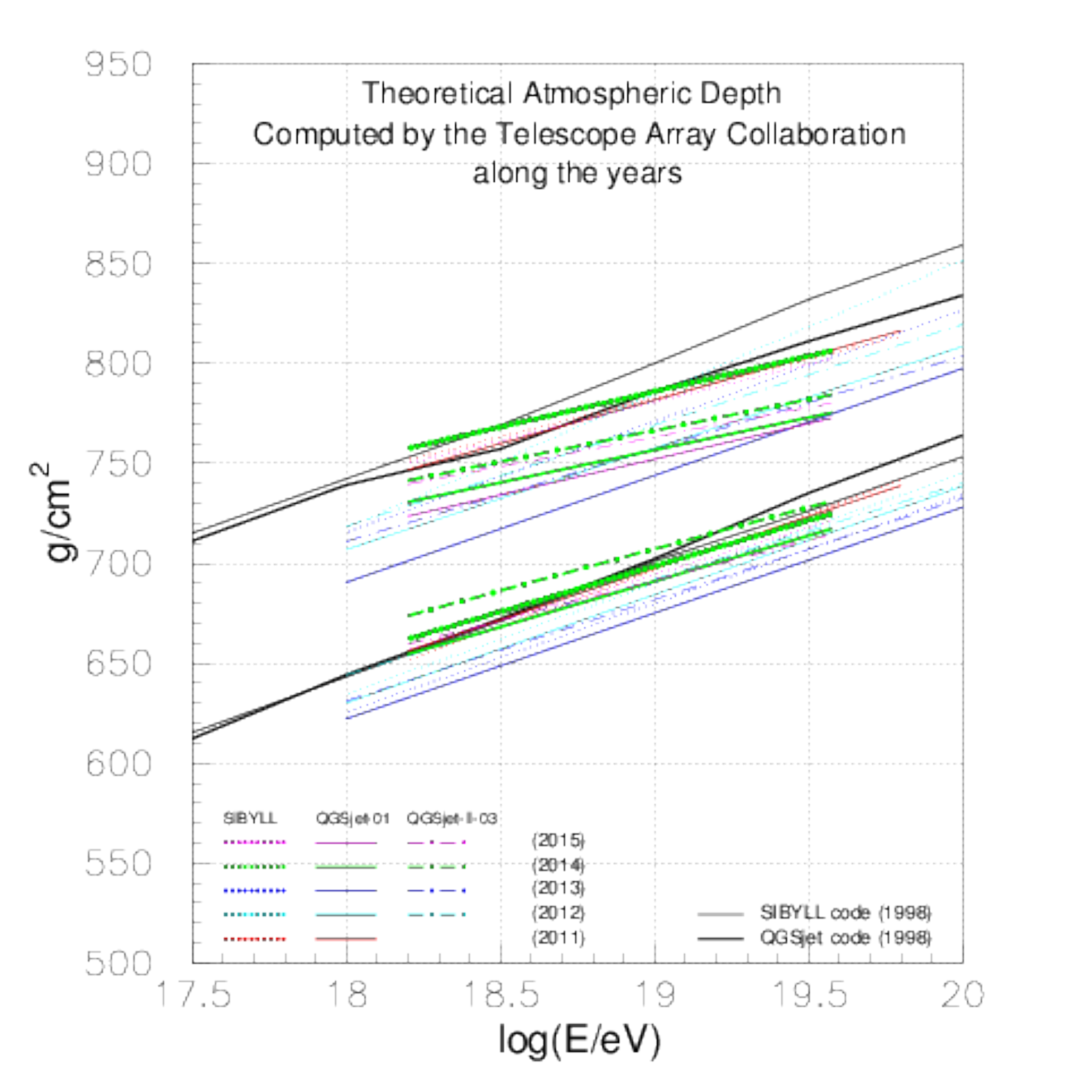}
    \caption{\label{fig_4}Mean atmospheric depth  X$_{max}$  versus energy computed  by the TA Group in  the years 2011-2015 and the hadronic simulation code used in the calculation.  The separation  between H and Fe  rails depends on the hadronic model and the energy. The absolute values  of  X$_{max}$   depend primarily  on the ensemble of proton-air cross sections.}
\end{center}
\end{figure}

In essential terms,  the TA and HiRes  Groups affirm the contrary of the assertion (A) e. g. almost  all cosmic rays are protons with no evolution of the chemical composition toward heavy nuclei. Explicit awareness of this assertion is documented in many places;  for instance: `` \dots to resolve outstanding differences in the interpretation of conflicting  X$_{max}$  data.'' ( William F. Hanlow, 2013 \cite{hanlow})   where X$_{max}$  data  refer to both Auger and TA data; on the same token: `` \dots A comparison with X$_{max}$  distribution with model simulations  (QGSjet-II-03),  we showed  the primary composition is consistent with 100\%  proton and inconsistent with 100\%  iron for energies 10$^{18.2}$  eV $<$ 10$^{19.2}$ eV.''   (Masaki  Fukushima,  2015 ref. \cite{fukushima}).

From these premises an unequivocal conclusion on the tendency of  the chemical composition is neither easy nor restful because the only florescence instrument which can compete with the exposures of Auger detector is that operated by the  TA  Group. As it will emerge in a moment the moot point hinges on the simulation codes of atmospheric cascades and not  on the  X$_{max}$  data  themselves.

The errors of the  X$_{max}$  in TA and Auger detectors are, respectively,  16.3 g/cm$^2$  \cite{abbasi14} and   20  g/cm$^2$ \cite{cazon,unger09b}. Event selection, event reconstruction,  atmospheric condition and  telescope  calibration are major sources of the systematic error.  Notice that  separation of the H-Fe rails of   80-95  g/cm$^2$  in the band 10$^{18}$ -10$^{19}$  eV  (see figure \ref{fig_3} and \ref{fig_4})   is  close to the systematic and statistical error, and hence,  only the tendency of the chemical composition versus  energy can be  reliably assessed.

      Figure \ref{fig_3}  shows  the atmospheric depth  X$_{max}$   versus  energy measured by the TA \cite{fukushima,abbasi14} and  Auger \cite{unger07,ahn} experiments  detecting  florescence light produced in giant air cascades. 

\begin{figure}[h]
  \begin{center}
    \includegraphics[width=0.6\textwidth]{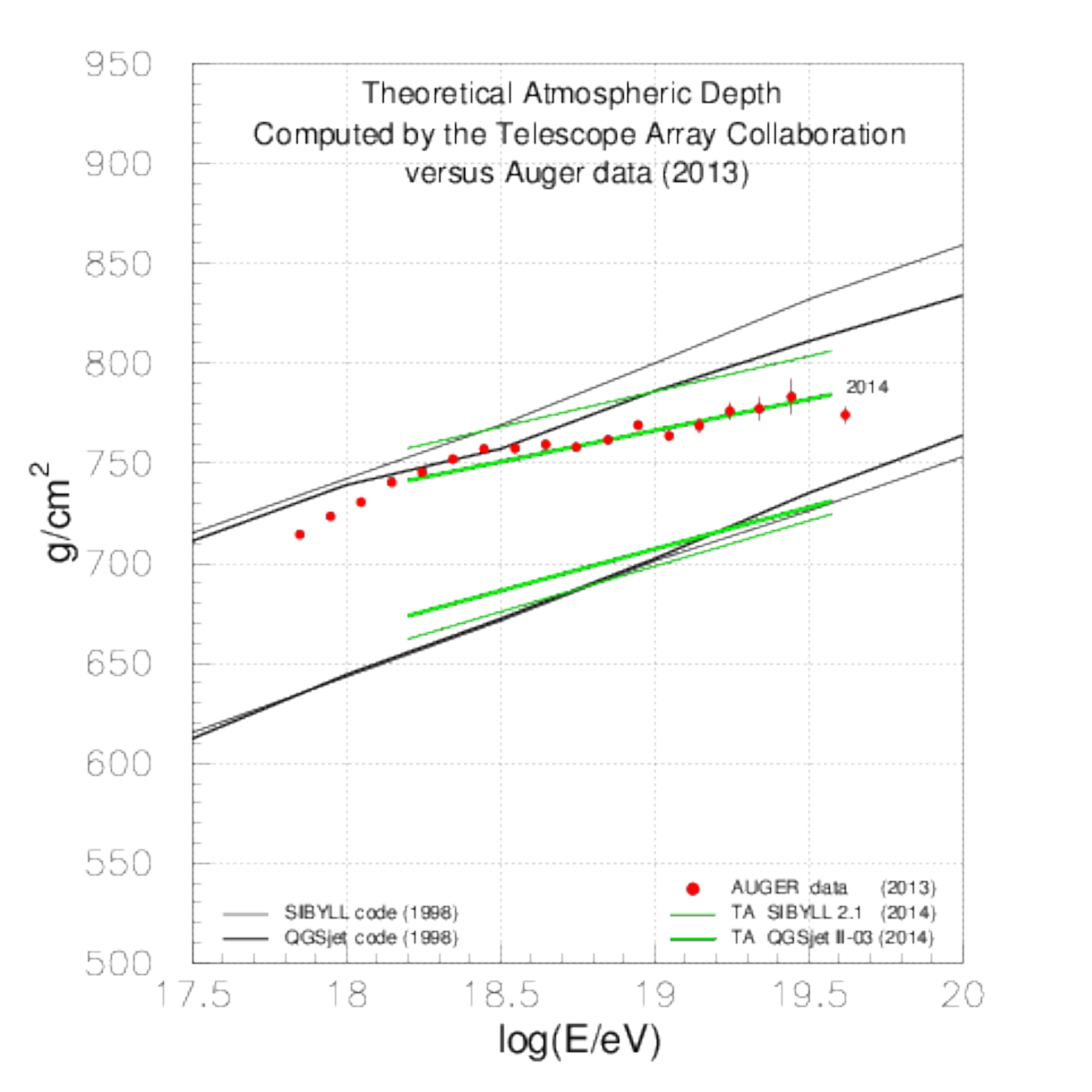}
    \caption{\label{fig_5}Measurements of  the  X$_{max}$  performed by the Auger experiment (red dots) \cite{ahn}  framed in the H and Fe atmospheric depths  (rails) evaluated by the  TA Group. The figure reports the classical   atmospheric depths  via  QGSjet and Sybill codes \cite{heck}  (black rails)  and those of the TA Group  (thick and  thin green rails).  Above 10$^{18.6}$ eV  data would be compatible with a purely  proton cosmic-ray component  except  the last data point.  The aim of  this figure is to draw attention to  the  ankle energy  band  e.g. 10$^{18.2}$ - 10$^{18.6}$ eV,  where  the Auger  X$_{max}$   data  are unphysical  if  the thick green rails   of  the QGSjet II-03  code  are  reliable calculations.   The thick green rail is labelled 2014  for clarity.}
\end{center}
\end{figure}
       
      The two rails (blue curves) in figure \ref{fig_3}  are theoretical atmospheric depths  for pure cosmic protons (upper rail) and pure cosmic Iron (lower rail)  and provide an aid to visually  and instantly reckon the chemical composition. These rails are classical predictions of atmospheric shower simulations \cite{heck}  that have been upgraded along the years, close and wide,  by new inputs on  nucleus-air cross sections, hadronic  fragmentation algorithms, inelasticities and  other variables.

      The abnormal lifting of the Auger  X$_{max}$  data  in the interval   10$^{17.9}$-10$^{19}$ eV  toward a light chemical composition  (see figure \ref{fig_3})  from year 2007  \cite{unger07}  to the year 2013 \cite{ahn} is embarrassing for a number of reasons.  The first one is that heavy ions cannot disappear from the spectrum  within the small energy interval 10$^{17.5}$ -10$^{18}$  eV. In fact many experiments around 10$^{17.5}$ eV reported a dominant fractions of heavy  nuclei including  HiRes-MIA  \cite{abu}  and HiRes  detectors \cite{abbasi05b}.  Secondly,  detailed calculations giving  nuclear species  versus  energy  in the interval 10$^{17}$ -10$^{19}$  eV thoroughly disagree with Auger data  of the year 2013  (see Figure 4 and 7 of ref. \cite{codino15b}).

Figure \ref{fig_4}  shows the bunch of  rails computed and adopted by the TA Group in the data analysis in the years 2011-2015. Patently, the computed TA atmospheric depths are so  scattered  that almost any interpretations of the experimental data in figure \ref{fig_3} become viable and legitimate.  As a vivid example,  consider the Auger data (red dots) \cite{ahn} and the  theoretical depth in Fig. \ref{fig_5} represented by  thick green rails.  The upper thick green rail (labelled 2014, for clarity)  is one of the many rails computed by the TA Group (also shown in fig. \ref{fig_4}) and it converts the X$_{max}$  Auger data in the range  10$^{19}$- 2.0$\times$10$^{20}$ eV  into purely  proton cosmic-ray composition. But if so,  around the ankle energy  e.g. 10$^{18.2}$ - 10$^{18.6}$ eV,   the Auger data  would become unphysical! Conversely, if the Auger data  (red dots) in fig.  \ref{fig_5}  are reliable  measurements,   then  the theoretical atmospheric depths are suspicious (thick green line 2014, via QGSjet-II-03).  A third possibility,  the most real one according to this and a previous study \cite{codino15b},  is that  both  Auger  X$_{max}$  data in the limited range 10$^{18.2}$ - 10$^{18.6}$ eV of the year 2013 \cite{ahn}   and  the QGSjet-II-03 outcomes  over the entire energy range  of fig.  \ref{fig_4},  are  essentially  incorrect and misleading. Notice that recent measurements of proton-air cross sections are below those  of the QGSjet-II-03  code.

The trend of increasing fractions of heavy nuclei with increasing energy in the band 4.0$\times$10$^{18}$- 10$^{20}$ eV is well substantiated by  both TA and Auger experiments as shown in figure \ref{fig_3} with the classical H and Fe rails \cite{heck}.  No empirical evidence discrediting  the H and Fe rails in figure \ref{fig_3} is known.  Progress and refinements in hadronic codes simulating  atmospheric showers did  take place but no upheaval regarding the main, critical  parameters emerged in recent years. For  example,  new data on  proton-air cross section  in the range 10$^{12}$ eV- 5$\times$10$^{18}$ eV  lie  on a straight line (see, for example,  figure \ref{fig_1} of ref. \cite{ulrich}). This feature  suggests that intimate substructures of hadrons smoothly coexist while colliding,  regardless of the energy, implying no bumps or dips in the cross sections, and plausibly, in the theoretical X$_{max}$  versus energy. Nevertheless,  deficiencies in the codes  remain (for example, muon deficit on the  ground of  about 25 per cent).
It is concluded that above 10$^{19}$ eV the assertion (A)  is  empirically founded,  not only by the   X$_{max}$  measurements shown in figure \ref{fig_3},  but also with the other two independent methods of measuring the chemical composition  \cite{cazon,unger09a} feasible with  the  Auger instrument.

\section{Premises of the spectrum calculation and their empirical basis}
The domestic origin of cosmic rays up to very high energy, 10$^{20}$  eV  and above, is not a predominant concept recurrent in the present and past literature \cite{codino15a}. As a consequence,  it is useful to  enumerate  those facts  suggesting that ultrahigh cosmic rays are galactic.  The first one  is the heavy composition of the cosmic radiation above 2.6$\times$10$^{19}$ eV. The second fact  is  based  on measurements of arrival directions of ultrahigh cosmic rays.

The evidence for the galactic  origin of   ultrahigh energy cosmic nuclei comes from  the nuclear photodisintegration cross sections $\sigma$($\gamma$,A) and  measured features of ubiquitous cosmic photons with  density $\rho$  of 411 particles/cm$^3$  and mean energy of  6.76$\times$10$^{-4}$ eV.   Important reactions are the ejection of one or two neutrons according to the reactions,    $\gamma$  A   $\rightarrow$   (A-1)  n   or    $\gamma$  A   $\rightarrow$   (A-2)  2 n   where  $\gamma$  represents the  ubiquitous photon, A  is the mass number of the cosmic nucleus and  n is the ejected neutron. In the laboratory energies  15-20 MeV  the cross section  $\sigma$($\gamma$,A)  has one or more  peaks and it is in the range  10$^{-25}$-10$^{-27}$ cm$^2$, sharply descending at higher energies.   The resulting characteristic  path L  of a long wandering extragalactic nucleus  is,   L   =  1/($\rho$ $\sigma$($\gamma$,A),  too tiny for a  cosmic world of gigaparsec size.

Had cosmic rays above  E$_L$ = 2.6$\times$10$^{19}$ eV  been extragalactic, a tight  correlation  between  backwardly extrapolated arrival directions and  locations of particular  celestial bodies (for example active galactic nuclei)  would have been discovered.  This  correlation  has never been detected  though cherished \cite{abraham} and expected.

No celestial bodies within 25 Mpc from the Earth,   believed to be  potential cosmic-ray sources,  intercept the backward extrapolated trajectories of the  most energetic cosmic rays as charted by many measurements with fairly good resolutions  (see,  for example,   ref. \cite{kristiansen}). This assertion becomes  highly constraining,  imperious, by the absence of a correlation between the direction of Virgo cluster of galaxies and arrival directions of ultrahigh cosmic rays.

For sake of  completeness,  another important hypothesis of the calculation,   designated  as  assumption (B) in the preceding paper \cite{codino17},  is mentioned.  It regards  the existence of the spectral break discovered by HiRes Collaboration in 2004 \cite{abbasi}.  Presently it has an undisputed, well assessed,  empirical evidence and no data scrutiny is necessary. 
\section{Concluding remarks}
As argued in Sections 3 and 4  the hypotheses of the calculation of the cosmic ray-spectrum above 10$^{19}$ eV  \cite{codino17}  are empirically  founded and dodge the dominant,  toxic   theoretical prejudice \cite{codino15a}.

The predicted spectrum shown in figure \ref{fig_1} and \ref{fig_2}  (tiny green squares) in the range  (2-9)$\times$10$^{19}$ eV  is comprised between the Auger flux (too low)  and the TA flux (too high).   Above 9$\times$10$^{19}$ eV  the Auger data (Figure \ref{fig_1}) lie below the predicted spectrum.  On the contrary the  Fly 's Eye  flux  shown in  2  (black dots) is above the predicted spectrum up to the maximum energy of  3$\times$10$^{20}$ eV.

According to this study, the existence of a fifth  stigma in the cosmic-ray spectrum  is correctly interpreted  assuming the existence of  a preferential selection  mechanism which sieves  quiescent particles in the interstellar medium prior to acceleration.  More precisely,  the mechanism is expected to operate in the cold interstellar territories surrounding  the H II regions inside the  O B star associations of the Milky Way Galaxy.  In  previous papers \cite{codino17,codino13} this filtering effect was termed  \textit{Lack of particle Injection to the Galactic Accelerator} (concisely,  \textsc{liga} effect) and the particular energy where the \textsc{liga} effect materializes  designated by  E$_L$.

In the near future a conclusive  validation  of the predicted spectrum  of  Fig. \ref{fig_1}, besides precise and reliable measurements of the fluxes,  could come from  the measurements of the chemical composition above 10$^{20}$ eV.   At the energy of  6.76$\times$10$^{20}$ eV  the atmospheric  depth  X$_{max}$  is  predicted  to lie  below the theoretical Fe  rail  (imagine fig. \ref{fig_3},  with  extrapolated rails up to 10$^{21}$ eV ) since no nucleus lighter than Iron composes the cosmic radiation  above this energy.

\end{document}